\documentclass[journal=jacsat,manuscript=article]{achemso}

\usepackage[version=3]{mhchem} 
\usepackage{subcaption} 
\usepackage{graphicx}
\usepackage{hyperref} 
\usepackage{amsmath}
\usepackage{booktabs}

\author{Jayesh Goswami}
\altaffiliation{Equal contribution}
\author{Shataneek Banerjee}
\altaffiliation{Equal contribution}
\author{Snigdhadev Chakraborty}
\author{Srestha Roy}
\author{Atanu Ghosh}
\author{Mrutyunjaya Rath}
\author{Agniva Das}
\author{Mukul Sagar}
\author{Krishna Kumari Swain}
\affiliation[Unknown University]
{Department of Physics, Quantum Centre of Excellence for Diamond and Emergent Materials (QuCenDiEM), IIT Madras, Chennai 600036, India}

\author{Prasanta Pal}
\affiliation[BigPharma]
{SHIOM LLC, Westborough, Massachusetts, USA, and RIHUB, Rhode Island, USA}

\author{Basudev Roy}
\email{basudev@iitm.ac.in}
\affiliation[Unknown University]
{Department of Chemistry, Unknown University, Unknown Town}

\title[An \textsf{achemso} demo]
  {A label-free sub-diffractive technique for 3D intracellular tomography using thermally induced convection currents}

\abbreviations{IR,NMR,UV}

\begin{document}


\begin{abstract}

Conventionally, 3-dimensional cellular tomography can be done with light sheet or multi-angle observations. Recently, a new technique was introduced where the cell was rotated using convection currents to visualize the outer periphery (Liu et al., Nano Lett., 2023, 23, 5148). However, the work falls short of actually observing intracellular objects like organelles etc. In this manuscript, we modify the technique by relying on computer vision algorithm called Contrast Limited Adaptive Histogram Equalisation (CLAHE) to improve the contrast for better detection of intra-cellular points, and then use optical flow detection technique to extract the in-plane speed of the point. This then is used to extract the vertical location, knowing that at the bottom part of the sphere, the point would be moving in one direction, close to the center there would be much less motion, while in the top portion of the sphere, the point would be moving in the reverse direction than the bottom. The velocity allows the exact localisation of the point in the vertical direction. This process allows for sub-diffractive intracellular tomography. This technique can further allow high-resolution detection of fluorescent molecules inside the cell also, when combined with convective flows.  

\end{abstract}


 \section{Introduction}
 
 The study and analysis of the three-dimensional (3D) characteristics of cells and organisms are of fundamental importance for understanding the structure and network \cite{monks1998three,wu2021control}. Conventionally, cellular imaging is done by taking multiple sequential images of a two-dimensional (2D) focal plane to form a focal stack. This technique gives high resolution along the lateral axis (x-axis), but leads to low resolution and missing information along the other axis, which is insufficient to visualize and resolve subcellular phenomena\cite{swoger2007multi,lidke2012advances,li2021deep}. Other methods include total internal reflection fluorescence\cite{mattheyses2010imaging,axelrod1983total,fish2009total,chung2007two,sako2002total} and selective plane illumination microscopy (SPIM) \cite{huisken2004optical,huisken2009selective,fu2016imaging,gao2015optimization,hedde20153d,huisken2007even} but this technique focuses on only a small portion of the sample close to the objective for high-resolution imaging\cite{axelrod2001total,power2017guide}. 
 
 Recently, multiview SPIM has also been introduced by adding multiple objective to scan the sample through multiple directions to overcome the detrimental effects of occlusion, scattering, and optical shadowing and improve the overall 3D imaging\cite{he2020image,chang2021real}. However, this technique also cannot provide full 360° optical imaging with high resolution\cite{liu2023high}. In another technique, rotating the sample continuously can provide a full 360° optical imaging even with standard optical microscope\cite{calisesi2020three,chhetri2015whole},by rotating the sample the missing information of the shadowing region can be captured. Thus, it can provide high-resolution imaging. However, this technique requires a fixed sample configuration on the substrate for stability and precision\cite{axelrod2001total}. The alternative way includes using external fields(i.e., optical, acoustic, and magnetic) for rotation. few such setups include optical and magnetic tweezers, the inclusion of these external field-based manipulation techniques is very helpful in various experiments to give best results as they are noncontact, compatible with biological samples, and easy to integrate with microfluidic devices\cite{ahmed2016rotational,tang2022rotation}. 
 
 Optical tweezers have been proven to provide precise control and generate good results in experiments involving cell size ranging from nanoscales to microscales. it can trap and can provide tomographic properties down to single-molecule level\cite{sudhakar2021germanium,moffitt2008recent}. optical tweezers uses tightly focused laser beams for trapping and performing 3D manipulation of micro-particles, cells, and proteins\cite{ndukaife2016long,pang2014optical,pang2012optical,marago2013optical,ashkin1987optical}. However, achieving stable rotation  with conventional optical tweezers is difficult due to the high mobility and asymmetry nature of the particles, which results in missing information and inaccurate optical imaging\cite{avsievich2020advancement,arzola2014rotation}. However, recent work reported that hydrodynamic flows provided an excellent and stable particle rotation \cite{panja2024nonlinear,nalupurackal2022hydro,sun2021rapid,junger2022100,schurmann2018three,nalupurackal2023controlled}. 

In this work we try to generate the cell rotation using thermophoretic flows. We create a sample chamber using a Dictyostelium cells as a sample, and putting them on gold-coated cover slip. As we direct the laser beam on the gold surface, it gets heated up and generates convection currents, which in turn lead to thermophoretic flows, inducing rotation of the cells. We extract the frames from the video of cell rotation. After extracting the frames, the contrast of each frame is enhanced using Contrast Limited Adaptive Histogram Equalization (CLAHE). The cell of interest is segmented out from the contrast enhanced frames. A number of intra-cellular points have been tracked from one frame to a subsequent one using Lucas Kanade Optical Flow. Finally the radial distance of those tracked points have been computed using a geometry-based method using properties of circle. The points plotted at the corresponding radial distances yields the three dimensional reconstruction of the cell.
 
\section{Theoretical details}

\begin{figure}[H]
    \centering
    \includegraphics[width=0.8\linewidth]{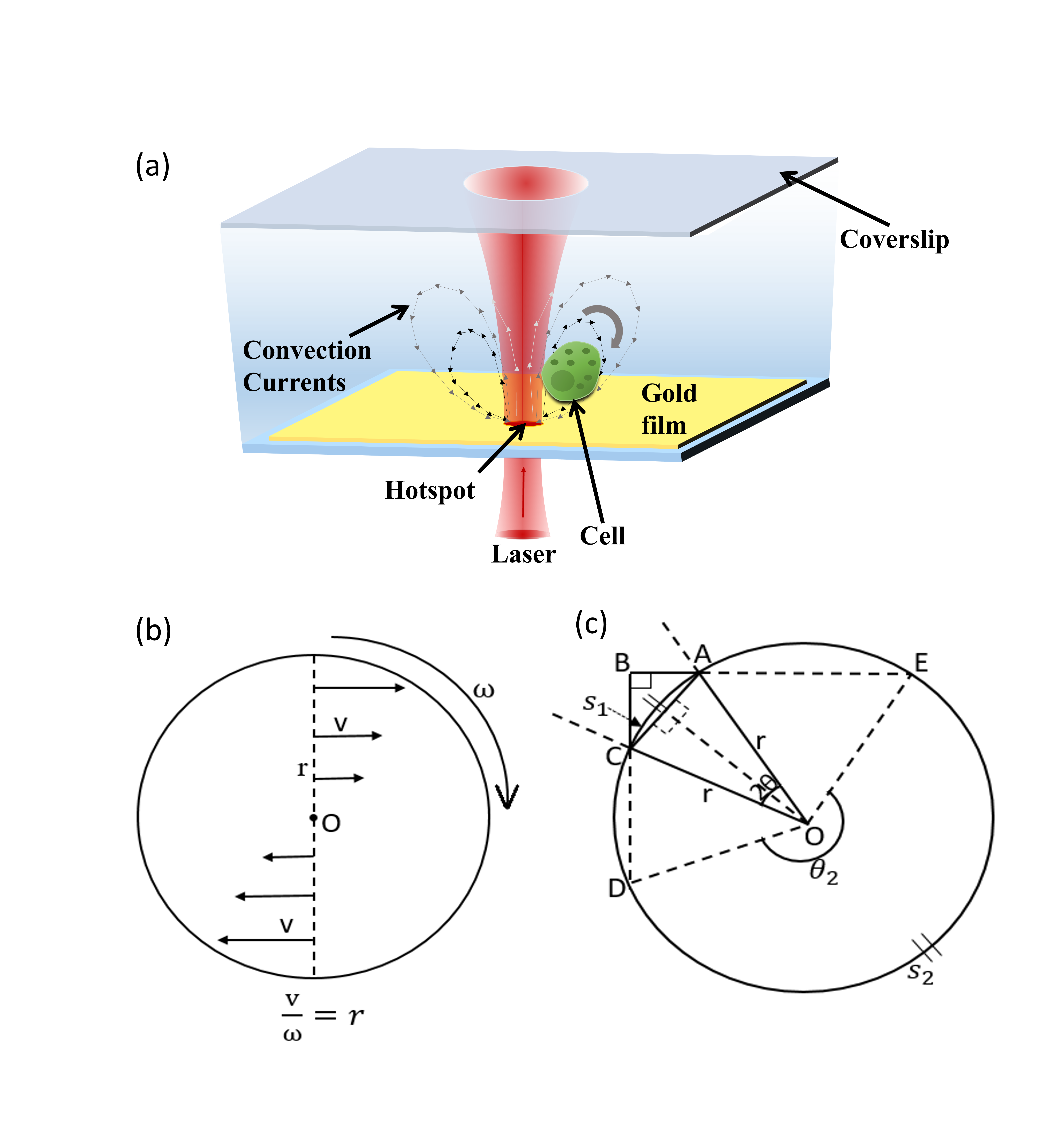}
    \caption{This figure shows the cell rotation mechanism and geometrical analaysis (a) shows the cell rotation under the infulence of convection current generated by the heating of gold film. (b) shows Intra-cellular velocity profile along the central line. (c) shows Calculation of radial distance of intra-cellular points.}
    \label{illustration}
\end{figure}

We generate the convection currents by using a  recently developed technique, where we first make a sample chamber with the bottom surface having a layer of gold on it. Then we illuminate a laser beam from the bottom objective into the bottom gold surface that induces convection currents close to the bottom surface. These convection currents then rotate the cell in the out of plane sense \cite{kumar2020pitch,liu2023high}, as shown in Fig. \ref{illustration} (a).

Once the cell rotates, the points inside the cell are made distinctly visible by enhancing the contrast using CLAHE and then their motion tracked. We extract the coordinates of the points, particularly the z location by a method shown in Fig. \ref{illustration} (b). Once the cell rotates, the outer points move with a velocity higher than the points towards the middle. Then towards the bottom, the points move reverse to the the motion at the top. However, in all of the cases of the points, the angular velocity is constant. This facet is used to first determine the angular velocity and then, from the velocity vectors, extract the depth at which the particle is located, as shown in Fig. \ref{illustration}(b) and (c). 

\begin{figure}[H]
    \centering
    \includegraphics[width=1\linewidth]{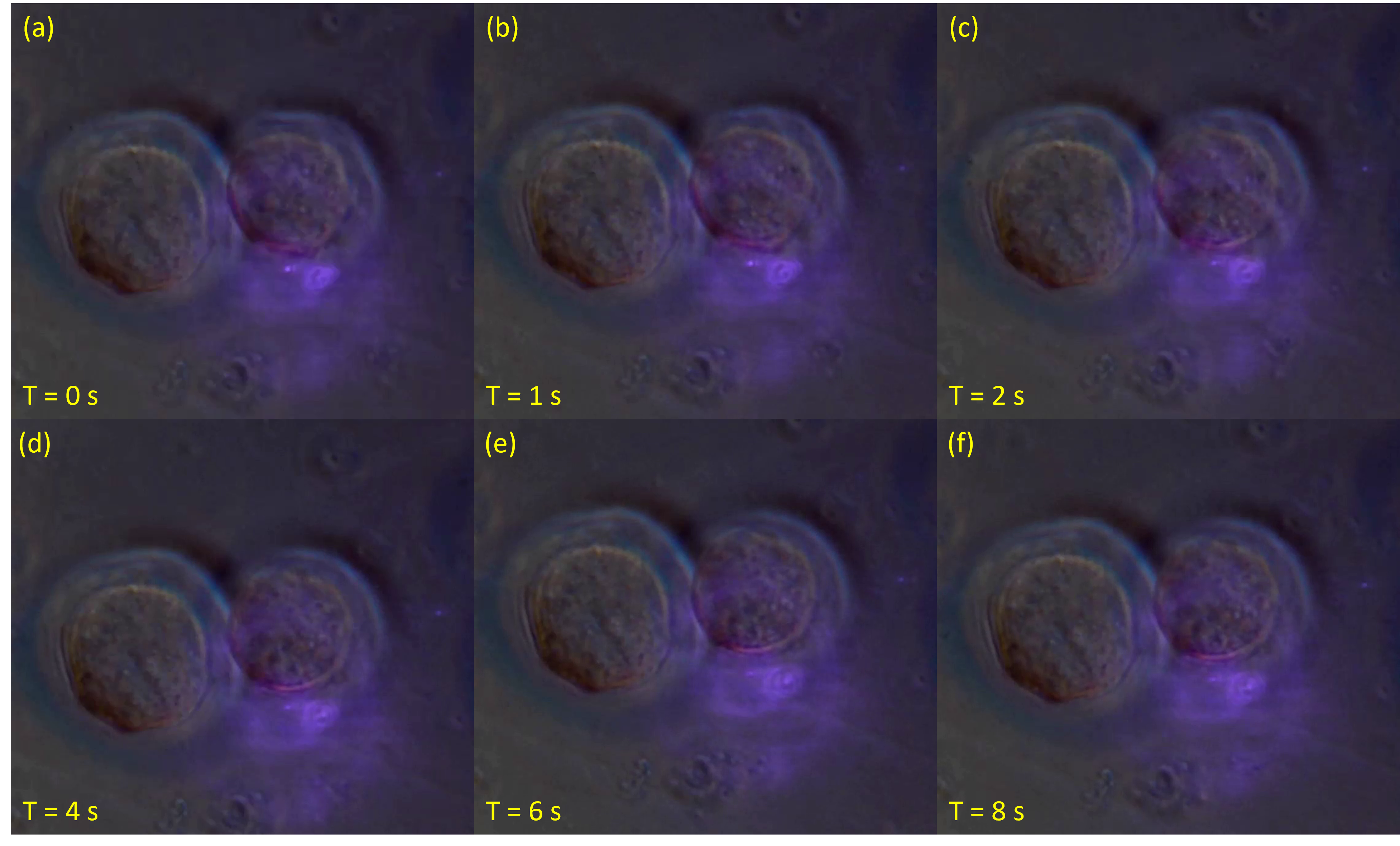}
    \caption{The figure shows the collage of diffrent snapshots of cell takan at diffrent time interval.}
    \label{video}
\end{figure}

In this work, we have tried to perform a 3D reconstruction  of a single cell interior using the technique mentioned. We have used Dictyostelium cells for our study. A typical event where the cell is rotated using the heated gold substrate has been shown in Fig. \ref{video}, taken with a phase contrast microscope at a very high Numerical Aperture (N.A.) of 1.49. This automatically assists in visualisation of sub-diffractive objects till about 100 nm or so.  

\subsection{Contrast enhancement of images}

\begin{figure}[H]
    \centering
    \begin{subfigure}{0.45\textwidth}
        \centering
        \includegraphics[width=\linewidth]{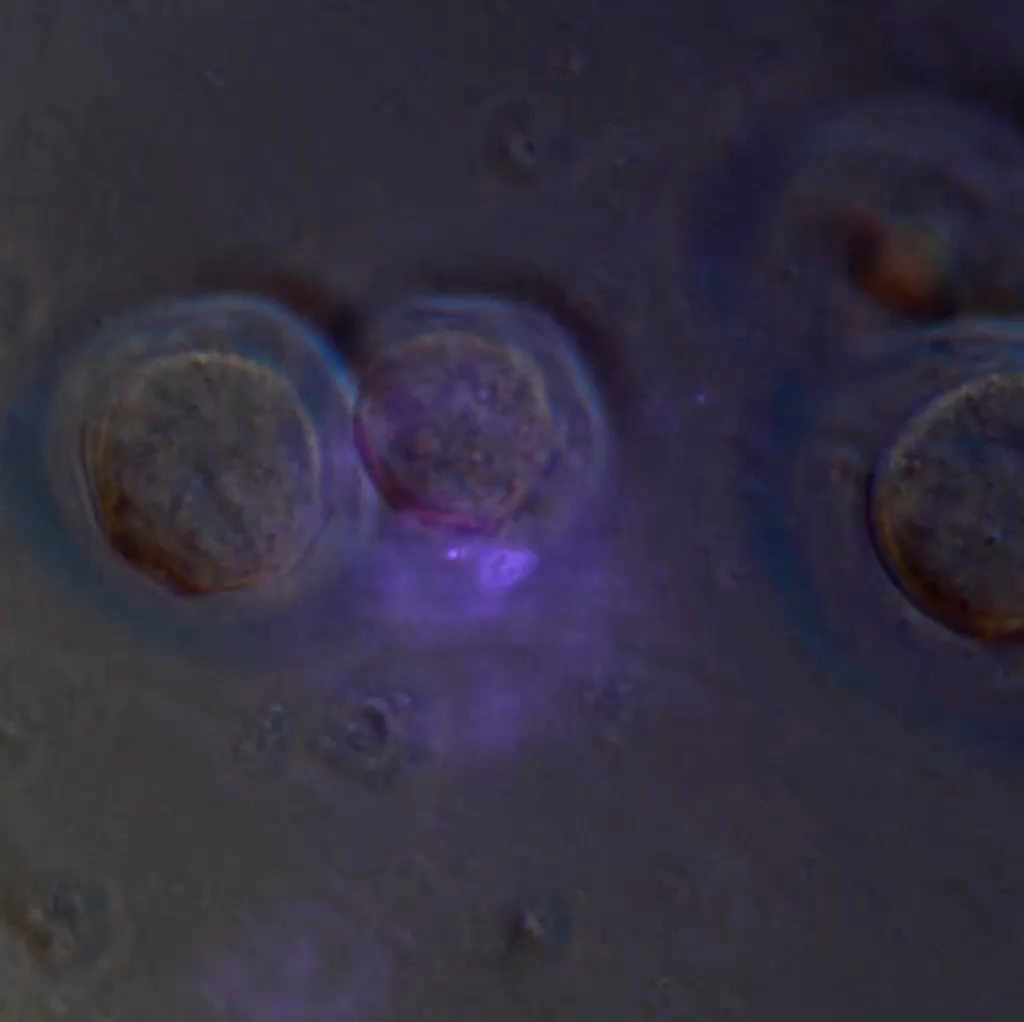}
        \caption{One snapshot of the video}
        \label{fig:subfig1}
    \end{subfigure}
    \hfill
    \begin{subfigure}{0.45\textwidth}
        \centering
        \includegraphics[width=\linewidth]{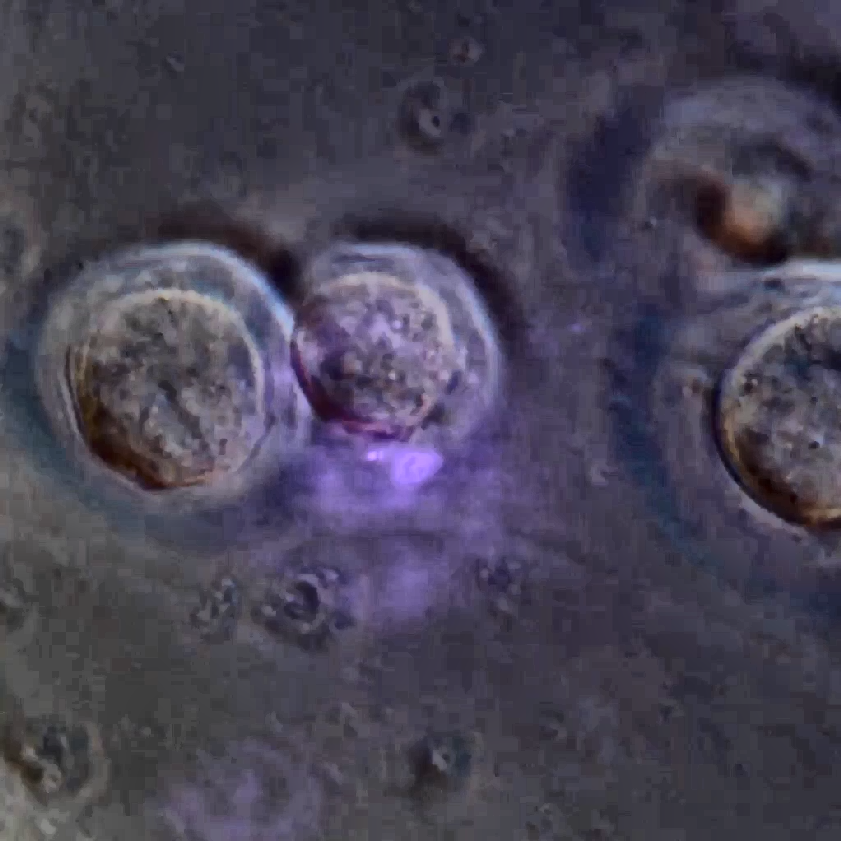}
        \caption{Contrast enhanced image}
        \label{fig:subfig2}
    \end{subfigure}
    
    \vspace{0.5cm} 

    \begin{subfigure}{0.45\textwidth}
        \centering
        \includegraphics[width=\linewidth]{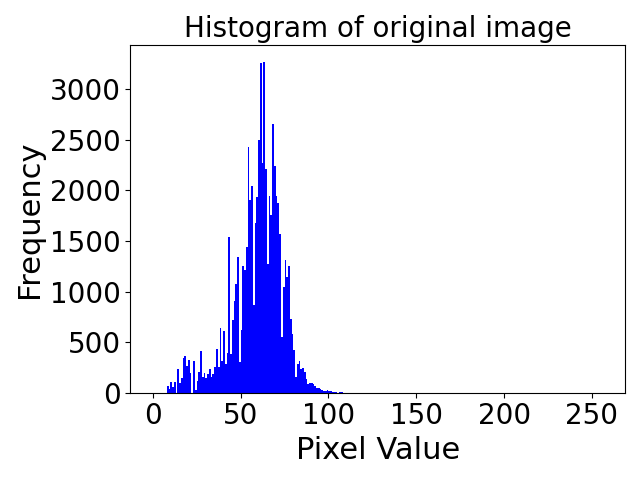}
        \caption{Histogram of original image}
        \label{fig:subfig3}
    \end{subfigure}
    \hfill
    \begin{subfigure}{0.45\textwidth}
        \centering
        \includegraphics[width=\linewidth]{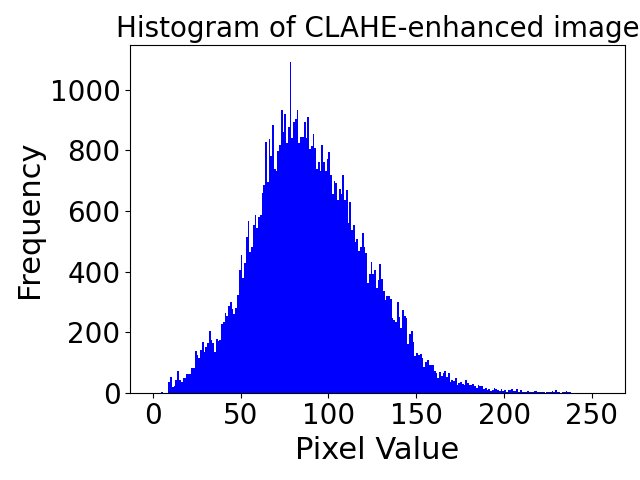}
        \caption{Histogram of contrast enhanced image. }
        \label{fig:subfig4}
    \end{subfigure}

    \caption{Application of CLAHE to images. Here we have increased the grayscale value of the pixels which are less frequent, while also decreasing the grayscale value of the pixels which have high frequency in the histogram}
    \label{fig:CLAHE}
\end{figure}

Different time snapshots inside the video have been enhanced using the Contrast Limited Adaptive Histogram Equalization (CLAHE) algorithm to facilitate
efficient detection of salient points within the cell. The image was divided into different sub-sections and a histogram (Fig. \ref{fig:subfig3}) was computed for each sub-section \cite{mondal2024single}. Then the pixel values were adjusted to increase the flatness of the histogram (Fig. \ref{fig:subfig4}). The way this was performed was to increase the grayscale value of the pixels which are less frequent, while also decreasing the grayscale value of the pixels which have high frequency in the histogram. This increase in flatness of the histogram leads to improvement of the image contrast and makes the points more distinctly visible \cite{yoshimi2024image}. 

The following equation demonstrates the application of CLAHE to an image\cite{chen2023contrast} :

\begin{equation}
I'(x,y) = \sum_{i=1}^{N} w_i \cdot T_i(I(x,y))
\end{equation}

where:

I(x,y) is the input pixel intensity at the location (x,y)

$I'(x,y)$ is the output pixel intensity at the location (x,y)

$T_i(I(x,y))$ is the transformation function for the i-th neighboring tile

$w_i$ are bilinear interpolation weights 

An example application of CLAHE to one snapshot of the video has been shown in Figure \ref{fig:CLAHE}.

\subsection{Calculation of radial distance of different intra-cellular points}

After the initial and final position of points have been tracked using optical flow, the radial distance of the points are calculated using a mathematical formulation that is based on rotational dynamics of spherical bodies. The equations to be solved have been listed below:

\begin{equation}
    v = \omega \times r
\end{equation}

\begin{equation}
    \frac{BC}{BD} = \frac{\sin(\theta)}{\sin(\theta_2/2)}
\end{equation}

\begin{equation}
    \frac{s_2 - s_1}{r} = \pi
\end{equation}

\begin{equation}
    CD = 2r \cos\left(\frac{\theta_2}{4} + \frac{\theta}{2}\right)
\end{equation}

The notations used in the equations are based on Figure \ref{illustration}(b) and (c). Solving the above equations for each point within the cell gives its corresponding radial distance.

\subsection{Point tracking using optical flow}

The cell of interest is segmented out from different snapshots of the video. Lucas Kanade (L-K) optical flow algorithm is applied on the segmented cell at two different time frames. L-K method is a widely used optical flow estimation method that uses the least squares criterion to solve the optical flow equations in a local neighborhood of a particular pixel.

L-K method solve the following system of equations\cite{al2023large}:

\begin{equation}
    A^T A v = A^T b
\end{equation}

where:

\( A = \begin{bmatrix} I_x(q_1) & I_y(q_1) \\ I_x(q_2) & I_y(q_2) \\ \vdots & \vdots \\ I_x(q_n) & I_y(q_n) \end{bmatrix} \)     \( v = \begin{bmatrix} V_x \\ V_y \end{bmatrix} \)     \(b = \begin{bmatrix} -I_t(q_1) \\ -I_t(q_2) \\ \vdots\\ -I_t(q_n) \end{bmatrix} \)

\begin{figure}[H]
    \centering
    \begin{subfigure}{0.45\textwidth}
        \centering
        \includegraphics[width=\linewidth]{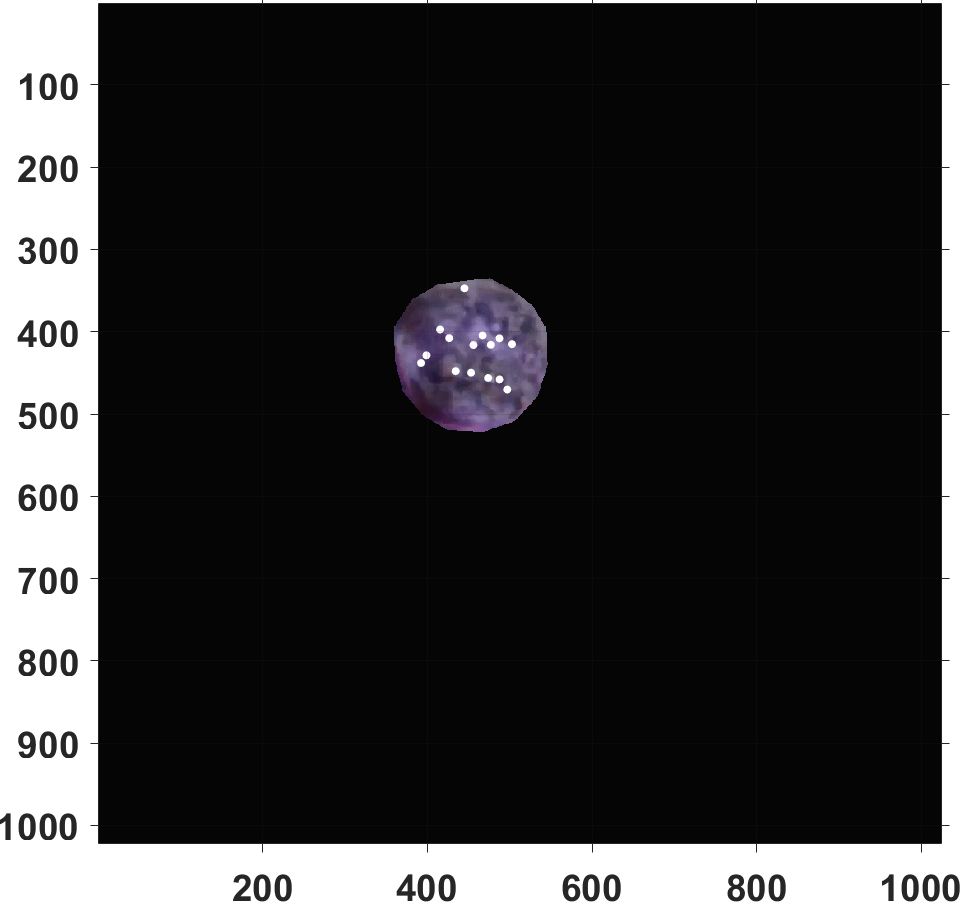}
        \caption{Position of points at one time frame}
        \label{fig:trackingsubfig1}
    \end{subfigure}
    \hfill
    \begin{subfigure}{0.45\textwidth}
        \centering
        \includegraphics[width=\linewidth]{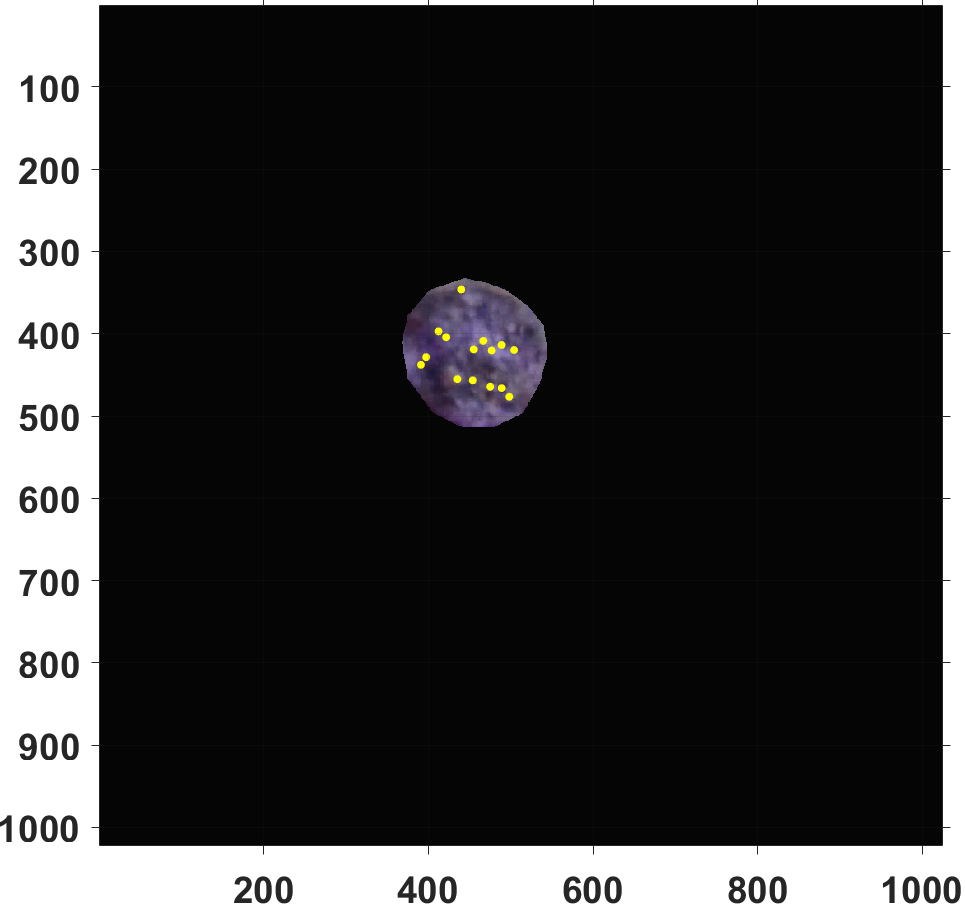}
        \caption{Position of points at a later time frame}
        \label{fig:trackingsubfig2}
    \end{subfigure}
    \caption{Point tracking by optical flow}
    \label{fig:Optical Flow}
\end{figure}

Optical flow estimation gives the positional shift of different spatial points inside the cell between the two time frames as shown in Figure \ref{fig:Optical Flow}. The coordinates of the points in the two frames are tabulated in Table \ref{tab:Coordinates}. The velocities of the points at the top and bottom of the cell are opposite in direction as demonstrated in the velocity profile shown in Figure \ref{illustration}.

\begin{table}[h]
    \centering
    \begin{tabular}{l c c c c}
        \toprule
        Point No. & \multicolumn{2}{c}{Earlier Time Frame} & \multicolumn{2}{c}{Later Time Frame} \\
        \midrule
         & x coordinate (pixels) & y coordinate (pixels) & x coordinate (pixels) & y coordinate (pixels) \\
        \midrule
        1  & 455.9  & 416.2  & 455.0  & 419.1  \\
        2  & 467.0  & 404.5  & 466.6  & 408.7  \\
        3  & 487.7  & 408.0  & 488.9  & 413.6  \\
        4  & 477.1  & 415.9  & 476.8  & 420.3  \\
        5  & 473.8  & 456.2  & 475.0  & 464.5  \\
        6  & 434.3  & 447.9  & 435.2  & 455.2  \\
        7  & 453.0  & 449.7  & 453.9  & 456.6  \\
        8  & 487.7  & 457.9  & 489.1  & 466.1  \\
        9  & 497.1  & 470.2  & 498.2  & 476.5  \\
        10 & 502.7  & 415.1  & 504.2  & 419.7  \\
        11 & 415.4  & 397.3  & 412.3  & 397.1  \\
        12 & 426.5  & 407.8  & 421.6  & 404.3  \\
        13 & 398.7  & 428.7  & 397.3  & 428.5  \\
        14 & 445.0  & 347.2  & 439.9  & 346.2  \\
        15 & 392.3  & 438.0  & 390.9  & 437.7  \\
        \bottomrule
    \end{tabular}
    \caption{Coordinates of different points in the earlier and later frames}
    \label{tab:Coordinates}
\end{table}

The fact that the minimum resolution of these coordinates is about 0.1 pixels, means that the resolution of detection of these points is about 10 nm, given that each pixels corresponds to about 100 nm. The localisation accuracy was further enhanced by enhancing the distinctiveness of the points using CLAHE. 


\section{Sample preparation}

Dictyostelium cells (wild-type D. discoideum AX2) were prepared by inoculating spores into 90 mm tissue culture plates filled with axenic HL5 growth medium (HLG01XX – Formedium, Norfolk, UK, pH 6.4), supplemented with  streptomycin sulfate (100 mg/ml) and penicillin (100 units per ml). The cultures were incubated at 22 °C until they became semi-confluent. Cells in the mid-log phase  were harvested in ice cold KK2 buffer (2.2 g/l KH2 PO4 and 0.7 g/l K2 HPO4 , pH 6.4), washed twice and 5 × 106 cells were pelleted and placed in ice for performing further tests.

\section{Experimental details}

\begin{figure}[H]
    \centering
    \includegraphics[width=1\linewidth]{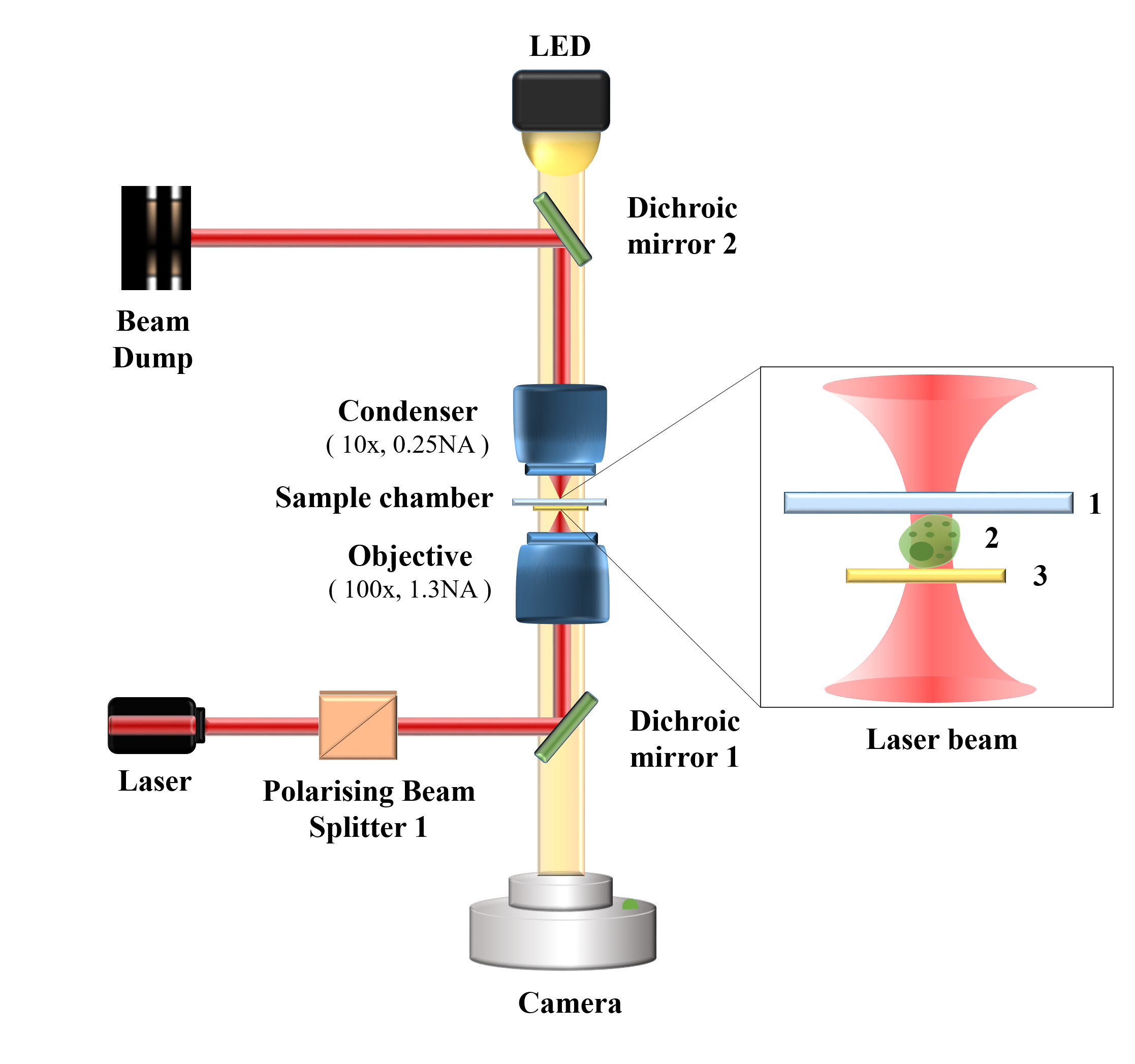}
    \caption{ The schematic diagram of the experimental setup is shown (1) Glass slide (2) Dictyostelium cells (3) Gold coverslip}
    \label{fig:setup}
\end{figure}

The experiment uses an inverted microscope (Nikon Eclipse Ti2E) integrated with an optical tweezers setup, equipped with a 60×, oil-immersion type objective lens with 1.49 Numerical aperture (N.A.) for simultaneous imaging and trapping. The condenser used was CLWD 10× with 0.72 N.A. (air-immersion) from Nikon. A 1064 nm wavelength solid state laser (Opus 1064) was employed for the experiment. The schematic representation of the experimental setup is depicted in Fig.\ref{fig:setup}.
The samples utilized in these experiments were Dictyostelium cells whose detailed preparation procedure is mentioned in the sample preparation section. A 20 $\mu$L aliquot of the sample was carefully placed on a glass slide (Blue star, number 1 size, English glass), then a gold-coated coverslip, (Blue star, number 1 size, English glass) of thickness 160$\mu$m with a 30 nm gold coating layer, was placed over the aqueous dispersion to complete the sample chamber. The prepared sample chamber was positioned on the sample stage of the optical tweezers setup, in an inverted position. A white light LED source is used to illuminate the sample stage. The output is collected by a CMOS camera placed at the output port for imaging purposes, as shown in Fig.1. The sample chamber was irradiated with a 1064nm laser beam, originating from the bottom of the chamber. Due to the plasmonic properties of gold, the regions where the laser beams interact with the gold-coated surface experience substantial heating, which in turn leads to an increase in the temperature of the surrounding medium, thereby inducing convection currents within the aqueous dispersion of cells. The temperature of the room was maintained at 25°C using an air conditioner.

\section{Results and Discussions}

\begin{figure}[H]
    \centering
    \includegraphics[width=0.7\linewidth]{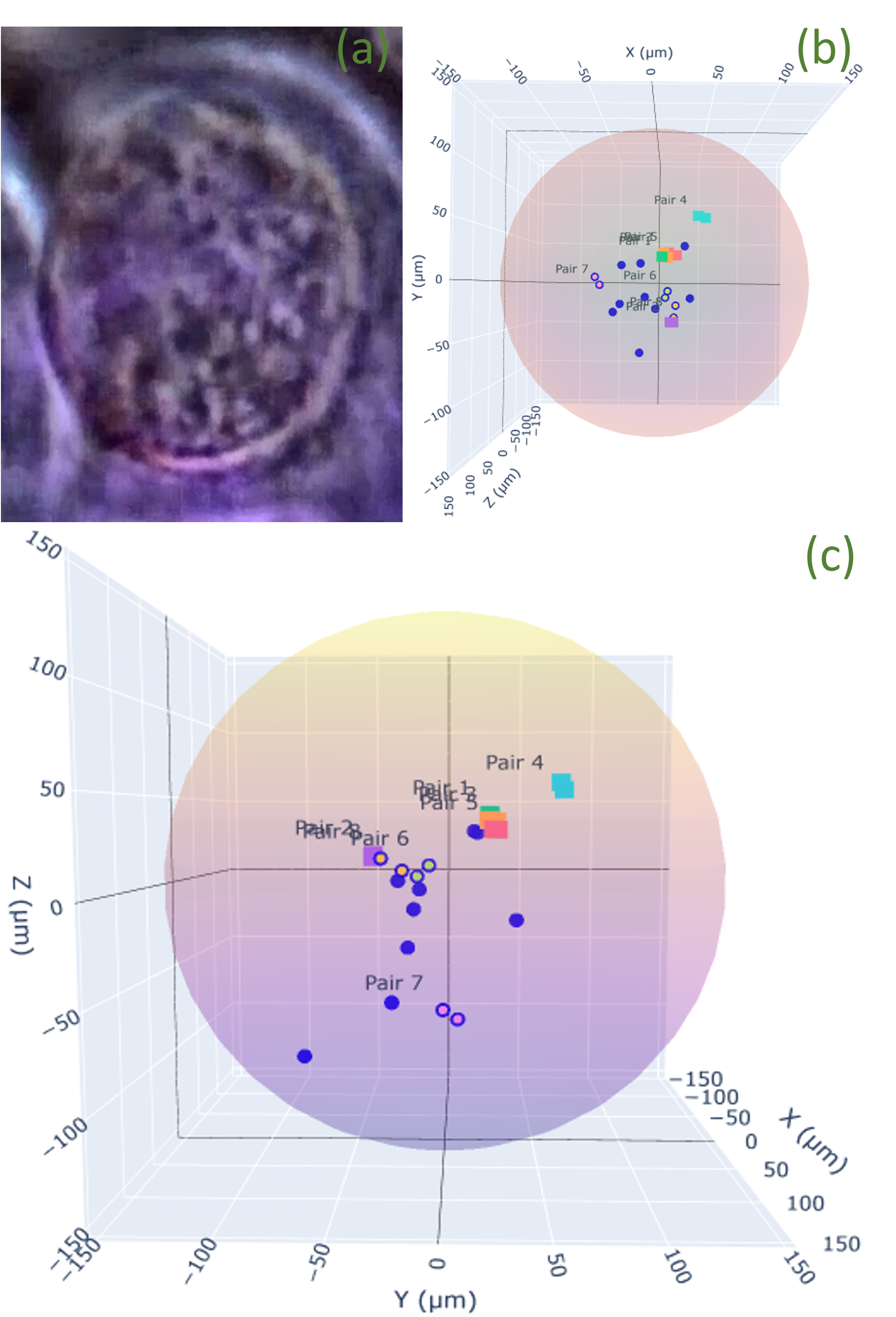}
    \caption{(a) Live cell image after significant contrast improvement, looked at from the top. (b) top view (c) front view The (a) and (b) may be different time points of the same event, but (b) and (c) correspond to same time point.}
    \label{fig:result}
\end{figure}

The 3-D map of the points inside the cell are made and the results depicted in Fig. \ref{fig:result}. The location of the objects inside the cell have been indicated from two different projections, and a bright field image shown for comparison. The points shown in the projections may have moved somewhat in the brightfield image, being from a different frame. It can be noticed that the resolution available for detecting these objects is limited by the resolution of detection of the x and y coordinates of the center of the spot. This is typically of the order of 10 nm. Thus, the resolution of detection of the z coordinate is also of that order and hence sub-diffractive. This bypasses the main problem in previous attempts where the spots could not be detected.

  In this manuscript, we use computer vision techniques to detect and track such objects inside the cell. Instead of just visualising the cell from different angles, we use a trick to locate the particles relying upon their speeds of motion, that allows sub-diffractive localisation of these. 

\section{Conclusions}

Thus, to conclude, we report a new technique to perform 3D intracellular tomography by rotating the cell partially and then use the velocity of the intracellular points as a signature of the depth of the particle. The method does not require complete 360 degree rotation of the cell, like used earlier, and yet performs the job of the tomography. The points are detected by using computer vision technique called CLAHE algorithm which is very good at extracting signals from noise and then using the optical flow detection technique to track the point over multiple frames. The method would assist conventional methods to detect even single fluorescent or label-free\cite{kasaian2024long} scattering objects to ascertain the vertical location of the point to sub-diffractive resolution of about 10 nm. 

\section{Conflict of Interest}

The authors declare that they have no known competing financial interests or personal relationships that could have appeared to influence the work reported in this paper. P.P. has co-founded a company, SHIOM LLC to commercialize technologies involving improvement in signal of noise of images.

\begin{acknowledgement}
We thank the Indian Institute of Technology, Madras, India, for their seed and initiation grants to Basudev Roy. This work was also supported by the DBT/Wellcome Trust India Alliance Fellowship IA/I/20/1/504900 awarded to Basudev Roy.

\end{acknowledgement}


\bibliography{achemso-demo}

\end{document}